# Charge-State Dependence of Electronic Excitations in keV Ions Transmitted Through Solids Probed by Ion-Photon Coincidence Measurements

Kevin Vomschee[1*], Radek Holeňák[2], Svenja Lohmann[2], Eleni Ntemou[1] and Daniel Primetzhofer[1,2]

[1]*Materials Physics, Department of Physics and Astronomy, Uppsala University, Uppsala, Sweden*
[2]*Tandem Laboratory, Uppsala University, Uppsala, Sweden*

## Abstract

Investigating the electronic excitations caused by light keV ions in matter contributes to a better understanding of materials modification such as space weathering or plasma-wall interactions in fusion reactors as well as improved materials analysis methods such as low or medium energy ion scattering. This study investigates the dependence of specific excitations in keV He projectiles, the most common projectile employed in these analytical methods, on the kinetic energy of the ion. We report on photon emission at energies >10eV emitted by projectiles excited upon transmission through different sample systems. We separate exit charge states and report on measurements of ion-photon pairs in coincidence, i.e. we can link a photon to the specific ion emitting it. From these measurements we extracted charge state resolved photon yields and obtained a deep insight into the electronic excitation occurring. Our results show that the electronic excitation of the helium projectile is largely material independent and rising strongly with the kinetic energy of the projectile. The photon yields are in good agreement with a common excitation model initially developed for beam foil spectroscopy. We hence extend the validity of this model to higher photon energies and different sample materials. We further provide a detailed analysis of the employed coincidence approach and necessary corrections to the raw data, as the methodology has a broad applicability in fundamental research and materials analysis.

*Corresponding author: kevin.vomschee@physics.uu.se

## 1. Introduction

Since the discovery of Rutherford Scattering [1], many Ion Scattering techniques have been developed and became reliable materials analysis tools. Many methods rely on measuring the energy loss of the backscattered primary particle in the interaction with the sample, among others these methods include Rutherford Backscattering (RBS), Medium Energy Ion Scattering (MEIS) [2] and Low Energy Ion Scattering (LEIS) [3]. For all these methods, He is the most commonly employed projectile, as it can backscatter elastically from any element with Z>2 while featuring a better mass resolution than H projectiles. The aforementioned methods achieve depth resolution due to the specific energy loss along the projectile path. This energy loss consists of two contributions: an electronic contribution where energy is deposited in the electronic system and a nuclear contribution where energy is transferred to the target atoms in multiple small angle collisions. While the scattering on target atoms and electrons is well described for the MeV energies used in RBS, reliable depth resolution on the nanometer scale requires keV energies as used in MEIS. A drawback in MEIS and LEIS is that the ongoing interactions are less well understood due to the lower projectile velocity, and that electronic and nuclear energy loss can even be entangled [4]. For setups using electrostatic or magnetic detectors electronic charge exchange is also of specific interest as only charged particles can be detected.

The same processes in keV ion-solid interaction are of relevance in plasma wall interactions within fusion reactors [5, 6] or in space weathering phenomena [7]; in both charge states of projectiles drastically alter the interaction. Investigating the photon emission of projectiles transmitted through a sample contributes to a better understanding of the electronic interactions. In the context of fusion reactors, charge exchange spectroscopy is even used as diagnostic tool investigating plasma properties [8, 9].

Historically, charge formation upon traversing matter was modelled as impact ionization ("electron stripping") [10, 11]. This picture is, however, oversimplified and today, empirical charge state formulas are commonly used [12]. For scattering experiments on collisions of single atoms in gases, more complex electron promotion models based on molecular orbital formation were shown to describe the interaction well [13]. Concerning solids, LEIS measurements utilizing large angle surface collisions identified similar electron promotion mechanisms (see [3] for a detailed review). However, while moving through the sample, projectiles commonly only experience small angle collisions, i.e. it remains unclear to which extend the aforementioned electron promotion processes known from gases and surface scattering contribute to the electronic energy loss in the bulk. Previous studies addressed these questions by measuring energy loss [14, 15, 16] and charge state distributions [17] across various crystal orientations. All these measurements show differences between "channeled" trajectories where projectiles travel along one of the crystals main axes and "random" trajectories where more close collisions are present (See [16] for more explanation). While both energy loss and charge state distributions point towards additional electronic excitations occurring in close collisions along random trajectories, exact processes remain a point of debate [18].

Electronic excitation mechanisms were also discussed for carbon targets in the context of beam foil spectroscopy [19, 20, 21] but the spectroscopy was limited to low energetic transitions, thus not ending in the ground state. Recently EUV photon emission upon keV ion transmission through solids was experimentally shown to be dominated by transitions occurring in the projectile behind the sample with several tens of percent of projectiles are leaving the sample excited [22]. EUV photons detectable with a microchannel plate (MCP) have >10eV energy so that the observed radiative transitions are for He limited to transitions ending in the ground state. While the observations in ref [22] gave an idea of the amount of excitation, the photons could not be linked to specific transitions or exit charge states. Utilizing the capability of a time-of-flight detection system to detect multiple, different signals in coincidence, in this contribution we investigate the kinetic energy dependence of charge specific EUV photon yields and hence get an idea of the excitation of every exiting projectile species (i.e. of He, $He^{+}$ and $He^{2+}$).

## 2. Experiment and coincidence treatment

As discussed in reference [22], tolerable beam currents on single crystalline samples do not allow for wavelength resolving EUV-beam foil spectroscopy. Accordingly, the photon origin cannot be deduced from the respective photon energies, our approach to assign individual photons to an exit charge state is hence based on Ion-photon coincidence.

Experiments were carried out using the Time-of-Flight Medium Energy Ion scattering (ToF-MEIS) setup at Uppsala University [23, 24]. The setup for charge state measurements is described in [17, 25]. Transmission experiments are performed with pulsed ion beams repeated at a frequency of 236 kHz with a pulse length of 0.5-3ns. The steady-state beam current is set such that in the vast majority of cycles at most a single ion is found in an individual pulse; thus, we can correlate projectiles and photons via coincidence conditions and obtain absolute photon yields of a specific charge. Transmitted projectiles and photons appear as distinct peaks in the recorded ToF spectrum [26]. The energy threshold for detection of photons with the RoentDek DLD120 delay line MCP-detector [27] is ca. 10eV approaching constant detection efficiency above 13eV [28, 29].

We performed coincidence measurements for $He^+$ projectiles incident on C foils as well as single crystalline Si and SiC membranes [30], with sample thicknesses of 50-200nm. For the Si case we distinguish between (100) axial channeling and a pseudo-random crystal orientation whereas for SiC only the pseudo-random case was recorded. Due to the fast charge equilibration and the low photon emission by the sample, the sample thickness and the initial projectile charge state are not expected to have a notable influence on our results. The transmitted beam is collimated by an aperture, and charge states are finally separated by an electric field. The charge-separated particles are detected by a RoentDek delay line detector [27] located 290mm behind the sample. While the horizontal angular interval of each charge is limited down to 1.6° by the aperture, we evaluate an 8° wide angular interval in the vertical dimension to increase statistics. This procedure is not expected to alter the obtained charge spectra. While charge state distributions were recorded as a side outcome of our measurements, we want to refer to [17] and [25] for distributions obtained on C and Si with the same setup.

The ion pulses have a length of 0.5-3ns (the temporal accuracy is however 0.5-10ns due to temporal shifts occurring during many hours of continuous operation). The repetition rate is $F \approx 236$ kHz while typical event rates are in the order of $\nu$=2-10 kHz and always held below 20 kHz. The detector is triggered with the same 236 kHz signal and starts a new time frame every time the trigger is recorded. A ToF spectrum is created by displaying events relative to the timing of the trigger (see ToF-MEIS description in ref. [24]). For He projectiles the obtained ToF spectrums show two main peaks: a first peak consisting of photons and a second peak consisting of transmitted projectiles. An exemplary ToF spectrum for ion transmission is e.g. shown in reference [22]. As the low event rate indicates, most of the time frames are empty and typically only ≈1-5% of the timeframes contain a single ion and its secondary particles. The coincidence condition can be formulated as follows: *"A count is coincident if within a timeframe the first event is registered within the photon peak and the second event is registered within the transmitted ion peak".* If this condition is fulfilled, we know which projectile created a specific photon and, if performed during a charge measurement, which charge state this projectile had.

As we measured the coincidence with different charge states, the photon detection is spatially limited to the Ω=7.17x10$^{-3}$ sr wide rectangular angular interval in which photons can pass the aperture of the charge deflector. The photon peak is well separated in the ToF spectrum as it is the peak temporally appearing first and a usually very narrow (0.5-10ns width) convolution of the jitter in the time domain of the actual ion pulse impacting the sample and the decay processes emitting photons. The decays of states with ns ranged lifetimes add up to a short exponentially decaying tail of the photon peak (see e.g. [26, 31, 22] for detailed descriptions of the photon peak). The temporal window in which the photon peak was evaluated was adjusted individually for each measurement, so that the exponential decaying tail was included. The resulting evaluated widths $\Delta t$ of the photon peak in the ToF-spectra were between 5ns and 10ns.

From the counts fulfilling the coincidence condition, absolute photon yields (i.e. photon emitting particle fractions) can be calculated if all following properties are considered (see [22]). The detection probability within the open area of the detector is $P_{ion,raw} = 1$ for ions and neutral atoms and largely constant at $P_{phot,raw} = 10\%$ for photons above 13 eV, while being 0 for photons below 10eV. The open area ratio of the detector is $P = 54\%$ i.e., the probabilities reduce to $P_{ion} = 54\%$ and $P_{phot} = 5.4\%$. For a spectrum with $n_i$ counts in charge state $i$ and $\tilde{n}_i$ counts of charge $i$ recorded in coincidence the absolute photon yield in photons/projectile is calculated as follows:

$$Y = \frac{\tilde{n}_i}{n_i} \cdot \frac{4\pi}{\Omega} \cdot \frac{1}{P_{phot}} \quad (1)$$

For wrong coincidence treatment we will also need a charge fraction $f_i = \frac{n_i}{\sum_i n_i}$, the overall number of coincidence events $\tilde{n} = \sum_i \tilde{n}_i$ and the overall number of photons in the measurement $n_{phot}$ after noise correction.

**Treatment of accidentally wrong coincidences**

Due to the strongly reduced probability of capturing photons through the charge state deflector, even after multiple hours of recording, the uncorrected value of $\tilde{n}_{raw,i}$ is commonly found only on the order of a few to several hundred counts. These counts contain also wrong coincidences and careful post processing is necessary. The contributions of wrong coincidences with noise $\tilde{n}_{noise,i}$ is proportional to the (non-coincident) charge state distribution. It can be reliably estimated by applying the coincidence condition to a 200ns wide noise area $\Delta t_{noise}$ in front of the photon peak where $\tilde{n}_{\mathrm{200ns_noise},i}$ projectiles will be coincident. The correction value $\tilde{n}_{\mathrm{noise},i}$ is usually not exceeding a few counts and can be expressed by:

$$\tilde{n}_{\mathrm{noise},i} = \frac{\Delta t}{\Delta t_{noise}} \tilde{n}_{\mathrm{200ns_noise},i} \quad (2)$$

The contribution of noise to the counts in the projectile time window is negligible so that no correction for noise below the projectile peak is necessary.

The largest source of wrong coincidences is caused by the presence of an ion that did not create the detected photon, the wrong counts generated by this are labeled $\tilde{n}_{wrong,i}$. Wrong coincidences of this kind are especially problematic if the divergence of the transmitted ion beam is large, as in this case many ions do not pass the aperture of the deflector but create detectable photons. The probability $P_{apert}$ describes the probability that an ion passes the aperture, this value is individual for each measurement. $\tilde{n}_{wrong,i}$ can also be well accounted for, the key size for this correction is the probability $P_{\mathrm{wrong}}$ that a wrong ion is detectable after photon detection. For the present small event rates $\nu$ this probability is well estimated by $P_{\mathrm{wrong}} = \nu/F$. The tree diagram clarifying all possible outcomes after photon detection is shown in fig. 1.

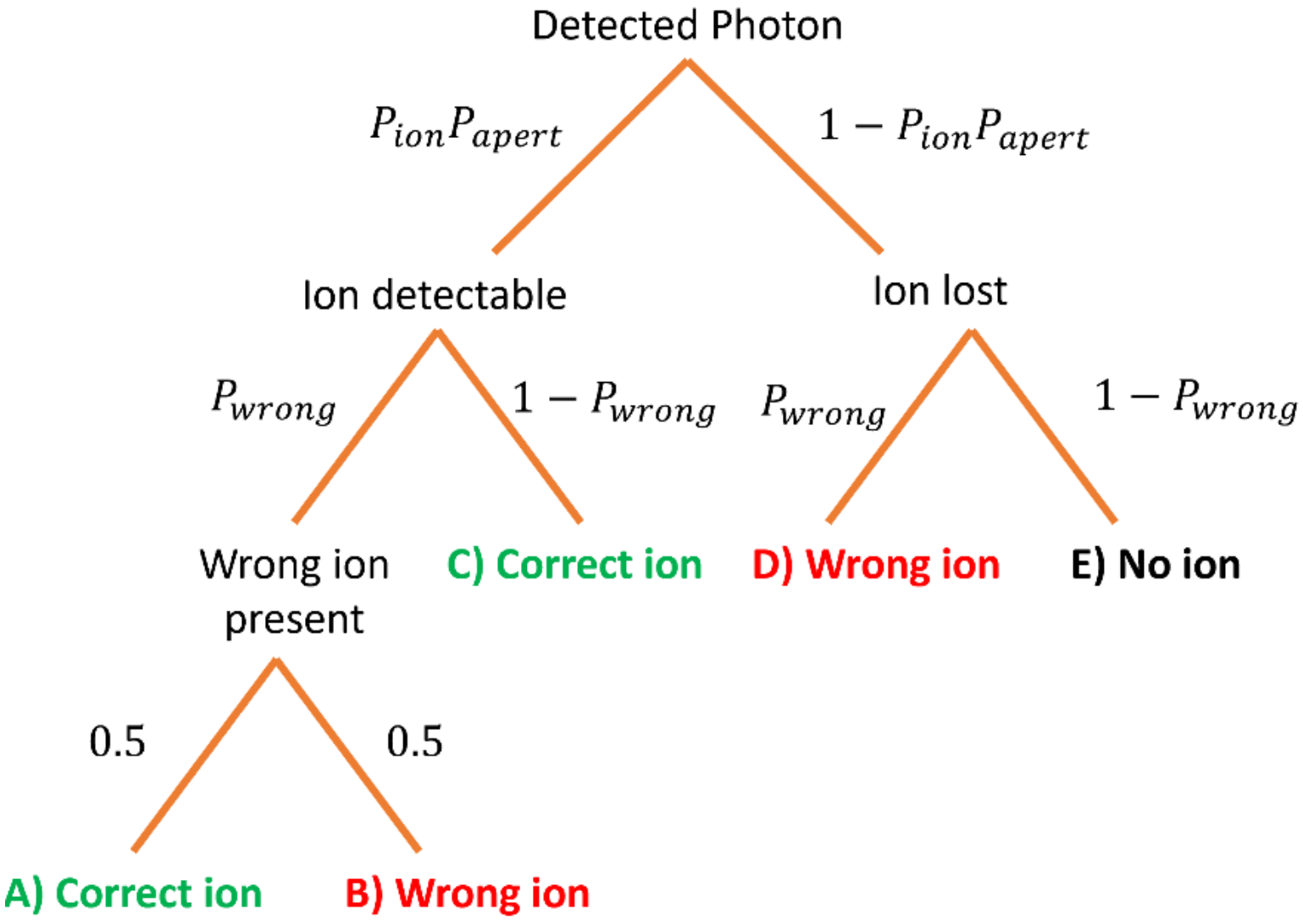

Figure 1: Tree diagram for the possible outcomes A)-E) after photon detection. A)-D) contribute to the raw coincidence data with the cases B) and D) causing wrong coincidences. The cases "ion detectable" and "Ion lost" refer to the correct ion. We ignore the case that two wrong ions are present which is parallel to case A) and B) but sufficiently unrealistic and in the order of $P_{wrong}^2$.

As shown in fig.1 there are five possible outcomes after photon detection with known probabilities. The wrong coincidences follow the non-coincident charge state distribution, so that after correcting for cases B) and D) we obtain the following expression for the real coincidences:

$$\tilde{n}_i = \frac{\tilde{n}_{\mathrm{raw,i}} - \tilde{n}_{\mathrm{noise},i} - f_i n_{\mathrm{phot}}\left(\frac{1}{2}P_{\mathrm{wrong}}P_{\mathrm{ion}}P_{\mathrm{apert}} + P_{\mathrm{wrong}}(1 - P_{\mathrm{ion}}P_{\mathrm{apert}})\right)}{1 - \frac{1}{2}P_{\mathrm{wrong}}} \quad (3)$$

The division by the term $1 - \frac{1}{2}P_{\mathrm{wrong}}$ is a correction emerging from the subtraction of case B). As in case B) the photon emitting ion was detectable, solely subtracting the ions from this case underestimates the photon yield and an upward correction of the remaining cases A) and C) is necessary. However, this upward correction is very minor and the majority of wrong coincidences is due to case D).

For applying eq. (3) it remains necessary to know the values of $\nu$ and $P_{apert}$ which are individual for each measurement. The value $\nu$ is the average event rate which is obtained by continuously recording the events accumulated within one second represented by $\nu_j$ being the number of events recorded in second $j$ after the start of the measurement. The average has to be done over all events so that we obtain:

$$\nu = \frac{\sum_j \nu_j{}^2}{\sum_j \nu_j} \quad (4)$$

The value $P_{apert}$ is not directly measurable but it can be iterated. The first iteration step is:

$$P_{\mathrm{apert}} = \frac{\sum_i \tilde{n}_{\mathrm{raw}} - \tilde{n}_{\mathrm{noise}}}{P n_{phot}} \quad (5)$$

Then $\tilde{n}_i$ is calculated with this $P_{\mathrm{apert}}$ and eq. (3) and the next iteration step is:

$$P_{apert} = \frac{\sum_i \tilde{n}_i}{P n_{phot}} \quad (6)$$

For a good iteration $P_{\text{apert}}$ and the following $\tilde{n}_i$ are calculated four more times. After this iteration the correction is finished and charge state resolved photon yields can be extracted according to eq. (1).

The relative error of the final result $Y$ is largely equivalent with the statistical error arising from $\tilde{n}_{\text{raw,i}}$, as $n_i$ is in any case >$10^5$ times larger than $\tilde{n}_{\text{raw,i}}$. Note that our measurements are largely unaffected by potential minor perturbations in the charge state distributions (e.g. from imperfect deflector placement) as such perturbations would only affect the correction (in $P_{\text{apert}}$) but not the ratio $\frac{\tilde{n}_i}{n_i}$ itself. Systematic errors arising from imperfect charge deflector placements have also only minor effects on $\Omega$.

# 3. Results

In Fig. 2 the photon emitting fraction of transmitted He projectiles as a function of the ion charge state and energy is shown for different target materials. The photon yields exhibit a strong energy (velocity) dependence and change drastically with the detected final charge state. Differences between C, Si and SiC targets are found small, despite differences of about a factor of 1.5 in electronic stopping power [32, 33] and the high atomic density of the SiC target. Axial ion channeling in Si does not impact the result, even though there is a demonstrated impact on electronic energy deposition [16]. In all measurements photon emission recorded in coincidence with detection of $He^{2+}$ is very low. This low photon yield of $He^{2+}$ is expected, as photon emission by the sample is very low [22] and no (potentially excited) electrons remain in the projectile. The photon yield for singly charged $He^{+}$ is about twice as large as the photon yield for neutrals. A linear velocity scaling, following the expected magnitude of electronic energy deposition [32] (square root behavior as indicated in fig. 2) describes the data satisfactory but the yields are found to drop more strongly at lower exit velocities (below the Bohr velocity $v_0$).

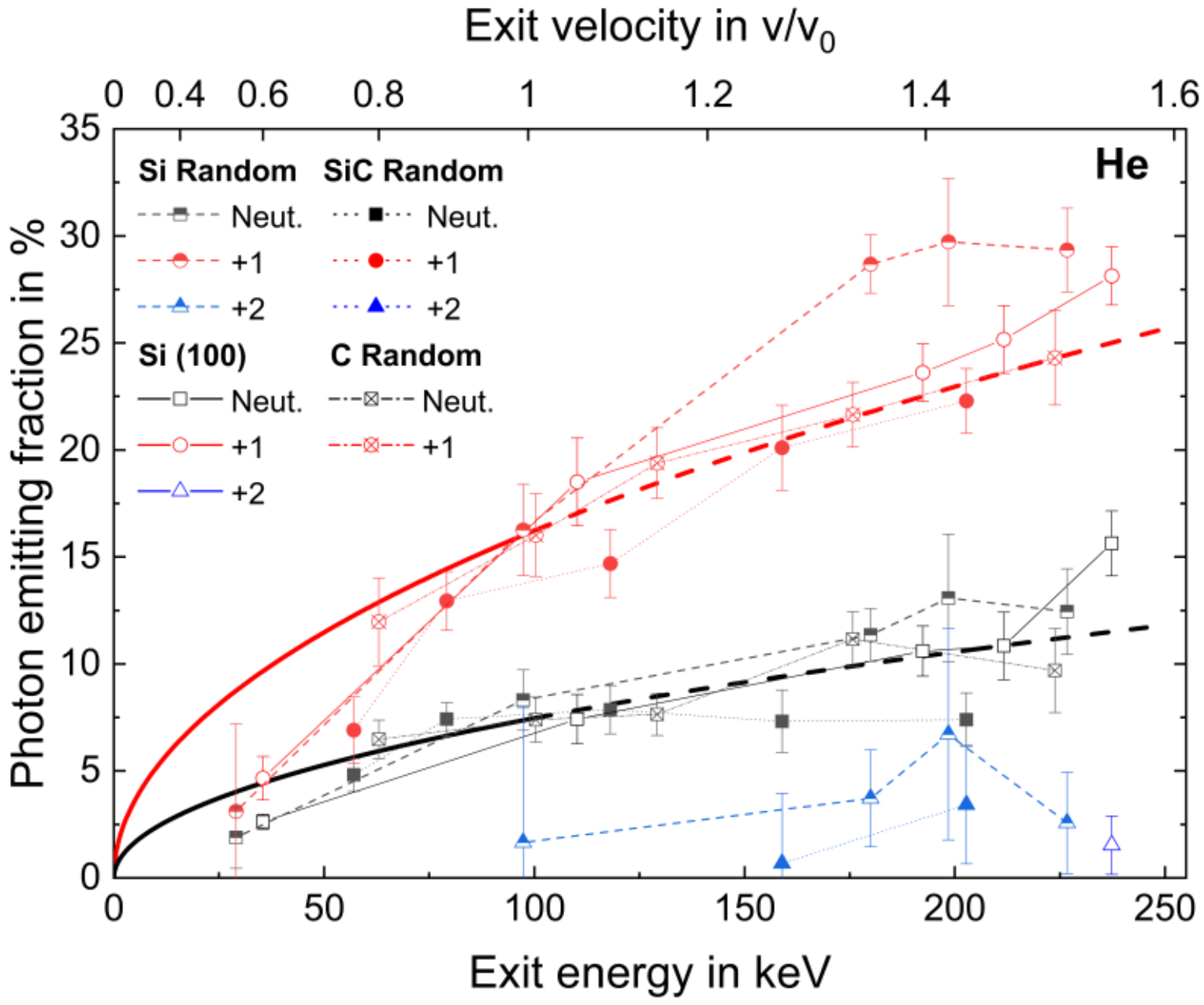

Figure 2: Photon emitting fractions of projectiles as a function of charge state after transmission for incident He+ projectiles, as a function of exit energy and velocity (top axis). The yields are plotted as photon emitting particle fractions. The thick solid/dashed lines display a square root behavior as guide to the eye; the different line type marks the Bohr velocity.

# 3. Discussion

Figure 2 shows a clearly distinct behavior for the three exit charge states. The low yields of $He^{2+}$ suffer from low statistics but give an idea of the comparably low photon emission emerging from the sample, i.e. the yield being dominated by the projectiles. Notably, these yields seem to be the lowest for the measurement in channeling geometry where Si core-excitation is expectably lower [16] and the measurement employing a carbon target, for which the absorption coefficients within the sample are high for <100eV photons [34].

The interpretation of the yields obtained for $He^{+}$ and neutrals is more complex. For any transitions in these freely decaying projectiles, we may define an initial main quantum number $n_i$ and a final main quantum number $n_f$ as well as a transition rate Γ [35]. Transitions need to have transition rates $\Gamma>10^8$Hz to contribute notably to the evaluated 10ns time window around the even narrower photon peak. For He, such transition rates can easily be compared to experimental values [36].

The detection efficiencies stated in section 2 suggest that we predominantly observe transitions into the 1s ground state (see [36] for transition energies). A few >10eV transitions ending in 2s and 2p exist for $He^{+}$ but have lifetimes of several ns, i.e. large contributions of these transitions would be visible in the ToF spectra and are not present. The results plotted in fig. 2 are hence free of cascades in which one ion could emit multiple photons. For He and $He^{+}$ we can therefore expect to primarily observe transitions of the kind $n_i$p -> 1s. With the restrictions for Γ we also know $2 \leq n_i \leq 5$ for Neutrals and $2 \leq n_i \leq 9$ for $He^{+}$ as visible transitions with higher $n_i$ exhibit too low transition rates. These initial states may be created by a faster, non-detectable, previous transition. We can furthermore assume, that excitations into 2p are most prominent, as the photon peak has an FWHM of only ~500ps in some of our measurements (even 360ps at 60keV in [26]) i.e. the peak is dominated by the width of the ion pulse while the comparably lower transition rates of deexcitations involving $n_i>2$ would cause broader peaks.

The observed dominance of $n_i=2$ is consistent with the $n_i^{-3}$ dependence of level populations reported in Beam Foil Spectroscopy (BFS) for radiative transitions with $n_f \geq 2$ [19, 20]. For 30-300 keV He in carbon, a model including only probabilities for loss of a 1s electron (probability $\beta$) and subsequent electron capture to an excited state (probability $\alpha$) was shown to reproduce charge state distributions and the level populations observed for $n_f \geq 2$ with good accuracy [20, 21]. If the model from reference [20] is reformulated for the fraction of excited $He^{+}$ ($He^{+*}$) we obtain:

$$\frac{P(\mathrm{He}^{+*})}{P(\mathrm{He}^{+*})+P(\mathrm{He}^{+})} = \frac{\alpha\beta}{1+(\alpha-1)\beta} \qquad (7)$$

Note for the interpretation of eq. (7), that electron capture to the ground state effectively manifests in a reduced $\beta$ as it is indistinguishable from no excitation occurring. The charge ratio between $He^{+}$ and $He^{2+}$ puts additional limitations to $\alpha$ and $\beta$ (see appendix). If the photon yield for $He^{+}$ and $He^{+}$ excitation are equivalent (i.e. if every excitation causes exactly one photon), we can extract the values for $\alpha$ and $\beta$ shown in figure 3.

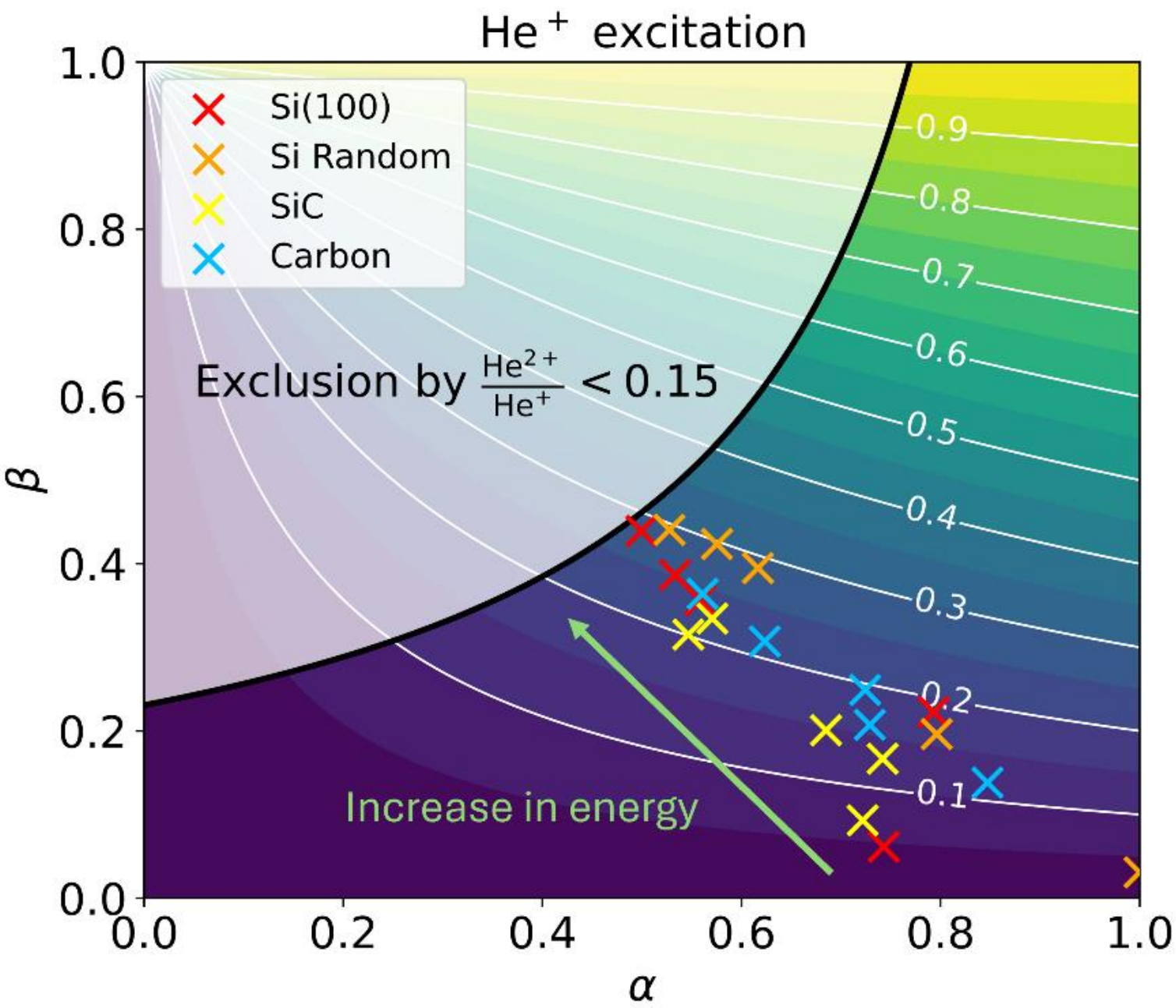

Figure 3: Equipotential lines display the fraction of excited $He^+$ in terms of α and β (see eq. (7)). With the additional limitations given the by $He^{2+}/He^+$ ratios values for α and β are extracted for all measured $He^+$ photon yields. The limitations given by $He^{2+}/He^+<0.15$ are shown as an example.

Figure 3 shows that the electron loss probability $\beta$ increases with increasing projectile energy whereas the capture probability $\alpha$ seems to decrease. The general behavior of $\alpha$ and $\beta$ is in good agreement with ref. [20]. Remarkably, there is no sharp drop in the capture probability $\alpha$ above the Bohr velocity: the photon yields keep increasing above the Bohr velocity despite the stripping criterion for excited electrons of He and $He^+$ being fulfilled as they have a maximum binding energy of 1Ry. Consequently, it should be much more likely to remove the 2p electron of $He^{+*}$ than to remove the 1s electron of $He^+$. Stripping of the excited electron would ultimately lead to the formation of $He^{2+}$ which is only rarely detected and was used as the reference to extract $\alpha$ as shown in fig.3. The high capture rates $\alpha$ suggest that the excited electron deexciting in the observed radiative transition is only captured at the very rear surface as further electronic interaction would strip the excited electron immediately. The instability of $He^{+*}$ within the solid (at least above the Bohr velocity) is additionally raising the question if the $He^{+*}$ and $He^{2+}$ detected behind the sample are potentially originating from similar excited states in the solid as both are formed from a projectile leaving the solid with two holes in the 1s shell.

Nevertheless, fig. 3 should only be used to discuss the qualitative behavior of $\alpha$ and $\beta$, as excitation and photon yields are not completely equivalent. As we report absolute photon yields for $n_f$=1 our data provides a reliable measure of the minimum excitation present in the transmitted projectile. However, some excitations will not emit a photon on the observed timescale, excitations into the 2s-state are for example extremely long lived for both He* and $He^{+*}$ if compared to the investigated 10ns time window. Additionally electronic deexcitation involving electrons from the sample (e.g. Interatomic Coulombic Decay [37, 38]) may deexcite excited projectiles nonradiatively [3]. Neutral He* is even more complicated as transitions are forbidden if both electrons occupy the same spin state. From BFS measurements for $n_f \geq 2$ a preference for triplet states and low angular momentum can be inferred [19], Triplet states are a likely explanation for the factor 2 difference in photon yields between neutral He and $He^+$. The photon yields stated in fig. 2 are hence only a lower limit of the overall excitation, while a proportionality between excitation and photon yields can be expected.

Within these limitations our data remains consistent in the combined capture and excitation model from reference [20]. The absence of significant differences in charge state distribution [17, 25] and photon yields for the different targets suggest that the findings of Davidson [19] and Dynefors et al. [20] apply universally. Consequently, Si, SiC and C targets exhibit nearly identical behavior in terms of excitation (i.e. electron loss) and subsequent electron capture into excited states and the population of these states expectably scales with $n_i^{-3}$. Even the specific states populated in the capture process must be very similar for all targets.

The overall agreement with a square root behavior depicted in fig.2 suggests that the photon yields scale with electronic energy loss to some degree as the electronic energy loss roughly follows a linear velocity scaling for the here investigated targets [32]. Nevertheless, the behavior at low projectile velocities shows a stronger drop in photon yield than expected from a linear velocity scaling. This drop can have various origins such as a lowered excitation probability or increased de-excitation via nonradiative Auger-like processes involving electrons in the sample that are also known to impact surface scattering [3]. That the excitation increases with projectile velocity is generally in good agreement with charge formation by processes like electron stripping. However, molecular orbital excitation for He would be non-adiabatic, and the excitation would therefore also exhibit strong velocity dependencies and potentially even a peak in the excitation probability [3, 39]. The exact velocity dependence of molecular orbital excitation would depend on target- and trajectory-specific interaction times and distances. As we find a monotonous increase in the excitation $\beta$, both molecular orbital promotion and electron stripping as well as a combination of these remain plausible excitation mechanisms. The absence of strong differences between the sample systems and sample alignments points towards the projectile excitation being dominated by electron stripping. Smaller, additional, contributions of molecular orbital promotion remain a likely explanation for the differences observed in electronic stopping [16], energy loss straggling [40] and charge state distributions [17].

So far, this discussion has ignored the case of $He^{**}$ formation. The high photon yields suggest a notable fraction of neutrals exiting in a doubly excited state with the total fraction being $\alpha^2\beta^2$. In many cases such a double excitation will be resolved by Auger electron emission. This Auger process will, dependent on the initial configuration, lead to the formation of either $He^+$ or $He^{+*}$. Therefore, the measured charge state distribution is not equivalent with the distribution upon leaving the sample and there is an additional impact on the $He^+$ photon yield. The formation of $He^{+*}$ would require two electrons to be captured at $n_i \geq 2$ which is rather unlikely. Hence the Auger decay of He** will predominantly form non-excited $He^+$, which is lowering the $He^+$ photon yield. Equation (7) and therefore fig. 3 ignores the contributions of He**. A better treatment requires knowledge of what fraction of He** projectiles decay radiatively, eq. (7) is equivalent with assuming complete radiative decay (i.e. no contribution to $He^+$). The other extreme (complete Auger decay of He** to non-excited $He^+$) is assumed in a separate calculation in the appendix, the calculated impact of He** on the extracted $\alpha$ and $\beta$ is very small.

# 4. Perspectives

Absolute quantification of photon yields by the presented ToF-MEIS coincidence technique has a range of applications in fundamental and analytical research. Measurements as performed here can be extended to more complex projectiles such as Ne or Ar where larger differences in charge state distributions [17] and stopping powers [15, 14] exist. The coincidence principle can also be used to separate the temporal coordinate of the detected photons by charge, which may give additional information when long lived transitions are present, as e.g. observed for transmission of Ne ions [22]. If the coincidence technique is applied in experimental setups with even shorter ion pulses [41, 42], exact initial states might be identifiable. Such measurements might allow better interpretations of theoretical studies e.g. based on Time Dependent Density Functional Theory (TD-DFT) [43, 44, 45].

# 5. Conclusion

Photon-projectile coincidence was applied to extract charge state resolved EUV photon yields of Helium ions transmitted through Si, C and SiC at various keV ion energies. The obtained charge resolved photon yields show similar energy dependencies as those known from less energetic transitions investigated in optical beam foil spectroscopy after transmission through carbon foils. An electron loss and capture based model developed for these datasets was shown to hold true for EUV-photons emitted from Si (including ion channeling) and SiC targets in transitions ending in the ground state.

Our results show a strong energy dependence of the excitation while electron capture to excited states retains a high likelihood across the whole energy range investigated. From the high amount of excitation and capture present, measurable Auger electron yields from He** are expected, these electrons perturbate the charge state distribution with respect to the distribution inside the solid. Capture to the more loosely bound (<1Ry binding energy) excited states remains high even at velocities where excitation of the deeply bound 1s electron is likely. This behavior suggests that the detected excitations are only formed at the very surface and that $He^{+*}$ and $He^{2+}$ originate from similar processes inside the solid.

The presented coincidence technique enables new kinds of measurements that will further improve the understanding of electronic interactions occurring in Ion-matter interaction.

# 6. Appendix

The ratio between the $He^{+}$ and $He^{2+}$ fractions can also be expressed in terms of $\alpha$ and $\beta$ since $He^{2+}$ is formed exactly then when $He^{+}$ loses an 1s electron without capturing it again. The relation used for calculating the points in fig. 3 is:

$$\frac{P(\mathrm{He}^{2+})}{P(\mathrm{He}^{+*}) + P(\mathrm{He}^{+})} = \frac{0.5\beta(1-\alpha)}{1+(\alpha-1)\beta} \qquad (8)$$

As stated in sec. 3, doubly excited He** is ignored in eq.(7) and (8) which could impact fig.3. Ignoring He** is here equivalent with assuming complete radiative decay, as this would only contribute to neutral He and its photon yield which are both are not part of eq. (7) and (8). The other extreme is complete auger deexcitation of He** to non-radiative $He^{+}$. For this eq. (7) and (8) have to be replaced with eq. (9) and (10).

$$\frac{P(\mathrm{He}^{+*})}{P(\mathrm{He}\,**) + P(\mathrm{He}^{+*}) + P(\mathrm{He}^{+})} = \frac{\alpha\beta}{1+(\alpha-1)\beta+0.5\frac{\alpha^2\beta^2}{1-\alpha}} \qquad (9)$$

$$\frac{P(\mathrm{He}^{2+})}{P(\mathrm{He}\,**) + P(\mathrm{He}^{+*}) + P(\mathrm{He}^{+})} = \frac{0.5\beta(1-\alpha)}{1+(\alpha-1)\beta+0.5\frac{\alpha^2\beta^2}{1-\alpha}} \qquad (10)$$

Figure 4 shows the results for $\alpha$ and $\beta$ if calculated based on eq. (9) and (10). While the functions behave very differently for both $\alpha$ and $\beta$ being large, the impact of He** on our dataset is rather small.

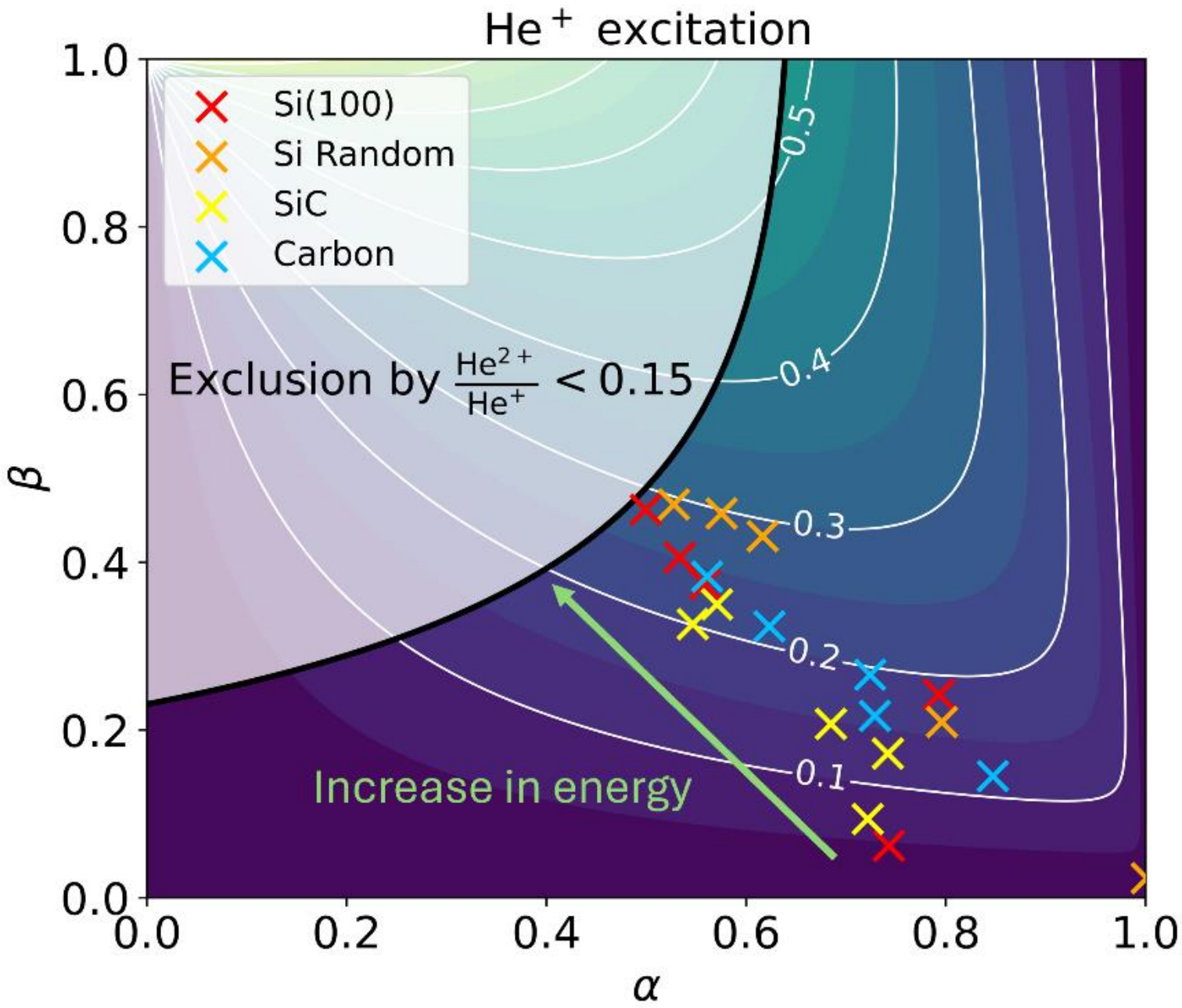

Figure 4: Equipotential lines display the fraction of excited $He^+$ in terms of α and β as deduced from eq. (9). With the additional limitations given the by $He^{2+}/He^+$ ratios and eq. (10) values for α and β are extracted for all measured $He^+$ photon yields. The limitations given by $He^{2+}/He^+<0.15$ are shown as an example.

# Acknowledgements

Accelerator operation is supported by the Swedish Research Council VR-RFI (Contract No. 2023-00155)

# Bibliography

[1] E. Rutherford, "LXXIX. The scattering of α and β particles by matter and the structure of the atom," *The London, Edinburgh, and Dublin Philosophical Magazine and Journal of Science,* vol. 21, p. 669–688, 1911.

[2] L. V. Goncharova, "Chapter 4.5.2 - Rutherford backscattering spectroscopy (RBS) and medium energy ion scattering (MEIS)," in *Characterization of Nanoparticles*, V. Hodoroaba, W. E. S. Unger and A. G. Shard, Eds., Elsevier, 2020, pp. 441-455. DOI: https://doi.org/10.1016/B978-0-12-814182-3.00023-7

[3] H. H. Brongersma, M. Draxler, M. de Ridder and P. Bauer, "Surface composition analysis by low-energy ion scattering," *Surface Science Reports,* vol. 62, pp. 63-109, 2007. DOI: https://doi.org/10.1016/j.surfrep.2006.12.002

[4] A. Niggas, L. Fischer, S. Kretschmer, M. Werl, H. Biber, C. Speckmann, N. McEvoy, J. Kotakoski, F. Aumayr, A. V. Krasheninnikov and R. A. Wilhelm, "Charge-exchange-dependent energy loss of

H and He in freestanding monolayers of graphene and $MoS_2$,," *Phys. Rev. A,* vol. 108, no. 6, p. 062823, 2023. DOI: https://doi.org/10.1103/PhysRevA.108.062823

[5] J. W. Coenen, M. Berger, M. J. Demkowicz, D. Matveev, A. Manhard, R. Neu, J. Riesch, B. Unterberg, M. Wirtz and C. Linsmeier, "Plasma-wall interaction of advanced materials," *Nuclear Materials and Energy,* vol. 12, pp. 307-312, 2017. DOI: https://doi.org/10.1016/j.nme.2016.10.008

[6] E. Pitthan, M. Fellinger, B. B. Domazet, P. M. Wolf, J. Shams-Latifi, F. Aumayr and D. Primetzhofer, "Interaction of light ions with plasma-facing materials: Improved experimental accuracy and its impact on sputter yield simulations," *Nuclear Materials and Energy,* vol. 45, p. 101996, 2025. DOI: https://doi.org/10.1016/j.nme.2025.101996

[7] P. S. Szabo, H. Biber, N. Jäggi, M. Brenner, D. Weichselbaum, A. Niggas, R. Stadlmayr, D. Primetzhofer, A. Nenning, A. Mutzke, M. Sauer, J. Fleig, A. Foelske-Schmitz, K. Mezger, H. Lammer, A. Galli, P. Wurz and F. Aumayr, "Dynamic Potential Sputtering of Lunar Analog Material by Solar Wind Ions," *The Astrophysical Journal,* vol. 891, p. 100, 2020. DOI: https://doi.org/10.3847/1538-4357/ab7008

[8] R. C. Isler, "A Review of Charge-Exchange Spectroscopy and Applications to Fusion Plasmas," *Physica Scripta,* vol. 35, p. 650, 1987. DOI: https://doi.org/10.1088/0031-8949/35/5/007

[9] O. P. Ford, L. Vanó, J. A. Alonso, J. Baldzuhn, M. N. A. Beurskens, C. Biedermann, S. A. Bozhenkov, G. Fuchert, B. Geiger, D. Hartmann, R. J. E. Jaspers, A. Kappatou, A. Langenberg, S. A. Lazerson, R. M. McDermott, P. McNeely, T. W. C. Neelis, N. A. Pablant, E. Pasch, N. Rust, R. Schroeder, E. R. Scott, H. M. Smith, T. Wegner, F. Kunkel, R. C. Wolf and W.-X. Team, "Charge exchange recombination spectroscopy at Wendelstein 7-X," *Review of Scientific Instruments,* vol. 91, p. 023507, 2020. DOI: https://doi.org/10.1063/1.5132936

[10] N. Bohr and J. Lindhard, "On the properties of a gas of charged particles," *K. Dan. Vidensk. Selsk. Mat.-Fys. Medd,* vol. 28, 1954.

[11] N. Bohr, "Scattering and Stopping of Fission Fragments," *Physical Review,* vol. 58, pp. 654-655, 1940. DOI: https://doi.org/10.1103/PhysRev.58.654

[12] G. Schiwietz and P. L. Grande, "Improved charge-state formulas," *Nuclear Instruments and Methods in Physics Research Section B: Beam Interactions with Materials and Atoms,* Vols. 175-177, pp. 125-131, 2001. DOI: https://doi.org/10.1016/S0168-583X(00)00583-8

[13] M. Barat and W. Lichten, "Extension of the Electron-Promotion Model to Asymmetric Atomic Collisions," *Phys. Rev. A,* vol. 6, no. 1, p. 211–229, 1972. DOI: https://doi.org/10.1103/PhysRevA.6.211

[14] F. H. Eisen, "Channeling of medium-mass ions through silicon," *Canadian Journal of Physics,* vol. 46, pp. 561-572, 1968. DOI: https://doi.org/10.1139/p68-070

[15] S. Lohmann, R. Holeňák and D. Primetzhofer, "Trajectory-dependent electronic excitations by light and heavy ions around and below the Bohr velocity," *Phys. Rev. A,* vol. 102, no. 6, p. 062803, 2020. DOI: https://doi.org/10.1103/PhysRevA.102.062803

[16] S. Lohmann and D. Primetzhofer, "Disparate Energy Scaling of Trajectory-Dependent Electronic Excitations for Slow Protons and He Ions," *Phys. Rev. Lett.,* vol. 124, no. 9, p. 096601, 2020. DOI: https://doi.org/10.1103/PhysRevLett.124.096601

[17] R. Holeňák, K. Vomschee, E. Ntemou, S. Lohmann and D. Primetzhofer, *Charge state dynamics of keV ions in solids,* 2026. DOI: https://doi.org/10.48550/arXiv.2412.01337

[18] F. Matias, P. L. Grande, M. Vos, P. Koval, N. E. Koval and N. R. Arista, "Nonlinear stopping effects of slow ions in a no-free-electron system: Titanium nitride," *Phys. Rev. A,* vol. 100, no. 3, p. 030701(R), 2019. DOI: https://doi.org/10.1103/PhysRevA.100.030701

[19] J. Davidson, "Relative and absolute level populations in beam-foil—excited neutral helium," *Phys. Rev. A,* vol. 12, no. 4, p. 1350–1357, 1975. DOI: https://doi.org/10.1103/PhysRevA.12.1350

[20] B. Dynefors, I. Martinson and E. Veje, "A Study of the Beam-Foil Excitation Mechanism Using 30-300 keV He+ Projectiles," *Physica scripta,* vol. 13, pp. 308-312, 1976. DOI: https://doi.org/10.1088/0031-8949/13/5/007

[21] E. Veje, "An independent-electron model for charge-state distributions and beam-foil populations," *Phys. Rev. A,* vol. 14, no. 6, p. 2077–2088, 1976. DOI: https://doi.org/10.1103/PhysRevA.14.2077

[22] K. Vomschee, R. Holeňák, C. Frank, S. Lohmann, E. Ntemou and D. Primetzhofer, *Deexcitations of the projectile dominate EUV photon emission upon transmission of keV ions through solids,* 2026. DOI: https://doi.org/10.1103/gt9t-rvn9

[23] M. A. Sortica, M. K. Linnarsson, D. Wessman, S. Lohmann and D. Primetzhofer, "A versatile time-of-flight medium-energy ion scattering setup using multiple delay-line detectors," *Nuclear Instruments and Methods in Physics Research Section B: Beam Interactions with Materials and Atoms,* vol. 463, pp. 16-20, 2020. DOI: https://doi.org/10.1016/j.nimb.2019.11.019

[24] M. K. Linnarsson, A. Hallén, J. Åström, D. Primetzhofer, S. Legendre and G. Possnert, "New beam line for time-of-flight medium energy ion scattering with large area position sensitive detector," *Review of Scientific Instruments,* vol. 83, p. 095107, 2012. DOI: https://doi.org/10.1063/1.4750195

[25] R. Holeňák, S. Lohmann, F. Sekula and D. Primetzhofer, "Simultaneous assessment of energy, charge state and angular distribution for medium energy ions interacting with ultra-thin self-supporting targets: A time-of-flight approach," *Vacuum,* vol. 185, p. 109988, 2021. DOI: https://doi.org/10.1016/j.vacuum.2020.109988

[26] E. Ntemou, R. Holeňák, D. Wessman and D. Primetzhofer, "Picosecond pulsed beams of light and heavy keV ions at the Time-of-Flight Medium energy ion scattering system at Uppsala University," *Nuclear Instruments and Methods in Physics Research Section B: Beam Interactions with Materials and Atoms,* vol. 556, p. 165494, 2024. DOI: https://doi.org/10.1016/j.nimb.2024.165494

[27] See http://www.roentdek.com/ for information on the delay line detector DLD120..

[28] R. C. Blase, R. R. Benke and K. S. Pickens, "Review of Measured Photon Detection Efficiencies of Microchannel Plates," *IEEE Transactions on Nuclear Science,* vol. 65, pp. 2839-2851, 2018. DOI: https://doi.org/10.1109/TNS.2018.2877356

[29] C. Martin and S. Bowyer, "Quantum efficiency of opaque CsI photocathodes with channel electron multiplier arrays in the extreme and far ultraviolet," *Appl. Opt.,* vol. 21, p. 4206–4207, 1982. DOI: https://doi.org/10.1364/AO.21.004206

[30] *Membranes manufactured by Norcada: https://www.norcada.com/.*

[31] S. Lohmann, M. A. Sortica, V. Paneta and D. Primetzhofer, "Analysis of photon emission induced by light and heavy ions in time-of-flight medium energy ion scattering," *Nuclear Instruments and Methods in Physics Research Section B: Beam Interactions with Materials and Atoms,* vol. 417, pp. 75-80, 2018. DOI: https://doi.org/10.1016/j.nimb.2017.08.005

[32] J. F. Ziegler, M. D. Ziegler and J. P. Biersack, "SRIM – The stopping and range of ions in matter," *Nuclear Instruments and Methods in Physics Research Section B: Beam Interactions with Materials and Atoms,* vol. 268, pp. 1818-1823, 2010. DOI: https://doi.org/10.1016/j.nimb.2010.02.091

[33] C. C. Montanari, P. Dimitriou, L. Marian, A. M. P. Mendez, J. P. Peralta and F. Bivort-Haiek, "The IAEA electronic stopping power database: Modernization, review, and analysis of the existing experimental data," *Nuclear Instruments and Methods in Physics Research Section B: Beam Interactions with Materials and Atoms,* vol. 551, p. 165336, 2024. DOI: https://doi.org/10.1016/j.nimb.2024.165336

[34] B. L. Henke, E. M. Gullikson and J. C. Davis, "X-Ray Interactions: Photoabsorption, Scattering, Transmission, and Reflection at E = 50-30,000 eV, Z = 1-92," *Atomic Data and Nuclear Data Tables,* vol. 54, pp. 181-342, 1993. DOI: https://doi.org/10.1006/adnd.1993.1013

[35] M. Werl, T. Koller, P. Haidegger, S. Wrathall, L. Eßletzbichler, A. Niggas, F. Aumayr, K. Tőkési and R. A. Wilhelm, "Lifetime of a freely decaying hollow atom," *Phys. Rev. Res.,* vol. 7, no. 1, p. 013176, 2025. DOI: https://doi.org/10.1103/PhysRevResearch.7.013176

[36] A. Kramida, Y. Ralchenko, J. Reader and N. A. S. D. Team, *NIST Atomic Spectra Database (ver. 5.8),* 2020. DOI: https://dx.doi.org/10.18434/T4W30F

[37] R. A. Wilhelm, E. Gruber, J. Schwestka, R. Kozubek, T. I. Madeira, J. P. Marques, J. Kobus, A. V. Krasheninnikov, M. Schleberger and F. Aumayr, "Interatomic Coulombic Decay: The Mechanism for Rapid Deexcitation of Hollow Atoms," *Physical Review Letters,* vol. 119, p. 103401, 2017. DOI: https://doi.org/10.1103/PhysRevLett.119.103401

[38] T. Jahnke, A. Czasch, M. S. Schöffler, S. Schössler, A. Knapp, M. Käsz, J. Titze, C. Wimmer, K. Kreidi, R. E. Grisenti, A. Staudte, O. Jagutzki, U. Hergenhahn, H. Schmidt-Böcking and R. Dörner, "Experimental Observation of Interatomic Coulombic Decay in Neon Dimers," *Phys. Rev. Lett.,* vol. 93, no. 16, p. 163401, 2004. DOI: https://doi.org/10.1103/PhysRevLett.93.163401

[39] S. Tsuneyuki and M. Tsukada, "Theory of the reionization process observed in low-energy $He^{+}$-surface scattering," *Phys. Rev. B,* vol. 34, no. 8, p. 5758–5768, 1986. DOI: https://doi.org/10.1103/PhysRevB.34.5758

[40] S. Lohmann, R. Holeňák, P. L. Grande and D. Primetzhofer, "Trajectory dependence of electronic energy-loss straggling at keV ion energies," *Phys. Rev. B,* vol. 107, no. 8, p. 085110, 2023. DOI: https://doi.org/10.1103/PhysRevB.107.085110

[41] L. Kalkhoff, A. Golombek, M. Schleberger, K. Sokolowski-Tinten, A. Wucher and L. Breuer, "Path to ion-based pump-probe experiments: Generation of 18 picosecond keV $Ne^+$ ion pulses from a cooled supersonic gas beam," *Phys. Rev. Res.,* vol. 5, no. 3, p. 033106, 2023. DOI: https://doi.org/10.1103/PhysRevResearch.5.033106

[42] A. Redl, M. Goldberger, C. Pachlinger, F. I. Pfanner and R. A. Wilhelm, "Generation of picosecond ion pulses using laser-stimulated desorption from a tungsten nanotip," *Phys. Rev. Res.,* vol. 7, no. 4, p. 043317, 2025. DOI: https://doi.org/10.1103/7wdx-97pf

[43] A. A. Correa, "Calculating electronic stopping power in materials from first principles," *Computational Materials Science,* vol. 150, pp. 291-303, 2018. DOI: https://doi.org/10.1016/j.commatsci.2018.03.064

[44] C.-W. Lee, J. A. Stewart, R. Dingreville, S. M. Foiles and A. Schleife, "Multiscale simulations of electron and ion dynamics in self-irradiated silicon," *Phys. Rev. B,* vol. 102, no. 2, p. 024107, 2020. DOI: https://doi.org/10.1103/PhysRevB.102.024107

[45] A. Lim, W. M. C. Foulkes, A. P. Horsfield, D. R. Mason, A. Schleife, E. W. Draeger and A. A. Correa, "Electron Elevator: Excitations across the Band Gap via a Dynamical Gap State," *Phys. Rev. Lett.,* vol. 116, no. 4, p. 043201, 2016. DOI: https://doi.org/10.1103/PhysRevLett.116.043201